# A REVIEW ON RECENT ACTIVE VIBRATION CONTROL TECHNIQUES

Gong Zuo,    Li Wong[1]

**ABSTRACT**

Active vibration control has been introduced and used as one of the effective approaches to suppress unwanted vibrations in different systems. Effective performance of each vibration control method is contingent to accurate design and proper dynamics selection of the control unit. These methods have been extensively studied in various studies in recent years. Each of these new methods are designed by a specific dynamics for a specific system. Here in this paper, we aim to introduce some of these recent approaches in a brief discussion, and familiarize the readers with these techniques. Engineers who wish to design proper vibration controllers in different scales, from micro- to macro applications, will certainly design a more successful vibration controller if they know better about similar techniques, and they can implement the novelties that other scholars have utilized.

[1] Corresponding Author, li.wong.me.eng@gmail.com

# 1. Introduction

Active vibration control has been used for rejecting the undesired vibrations in different systems for many years. The problem of undesired vibrations arise from an intrinsic problem in flexible structures that these systems are easily vibrated due to the task that they are assigned for, or due to severe ambient conditions. This problem is not limited to one system or one design, and a wide variety of systems have suffered from this issue. This problem happens when resonant modes of a piezoelectric stage are excited when scanning, or when a robot arm is moving under discontinuous forces at its end, or when a drone is being influenced by the wind in a severe weather.

The key point in designing a successful controller design is first to understand the problem very well. When the system is studied and analyzed completely, the source of the disturbance is known and the model of the system is extracted, the engineer needs to find the proper place that an actuator can be set, and the way that the feedback can be collected. Having a feedback from the vibrating structure is essential in designing the active vibration controller. Thirdly, they type of available and accessible sensors and actuators should be specified.

Having all these steps taken, it is the controller that lastly plays the most important role. The controller which is also known as the software of our system, can maximize the performance which can be obtained from the hardware of the system. To this end in this work, we present a brief review of the recent publications in this field. For each technique, we provide a brief discussion and provide an overview of the design in some cases.

## 2. INTEGRAL-BASED CONTROLLERS

Integral controllers are known as one of the most successful techniques for vibration control. These methods have a first order integrator, which increase the damping of the system when they are used in a closed-loop form. In a study by Szabat and Teresa [1], an analysis of control structures for the electrical drive system with elastic joint is conducted. They have used a proportional-integral controller supported by different additional feedbacks. A method for a robust integral controller is presented by Hu [2], and problem of pulse-width pulse-frequency modulated input shaper for flexible spacecraft has been completely discussed in this work. Integral twist actuation of helicopter rotor blades have also been used for vibration control in [3]. A series of vibration control methods are developed based on PID control method. PID such as its numerous applications in different control design systems, they have also been used for vibration controller. In these systems however, the integrator plays a very important role. Some very useful examples of these techniques are found at [4-7].

## 3. SLIDING MODE CONTROL AND NONLINEAR METHODS

One of the most frequently used approaches for vibration control is Sliding Mode Control or (SMC). An adaptive method is proposed and experimentally used by Li et al [8]. Another method has been developed by Hu which is an observer based method [9].

A series of nonlinear vibration controller have been implemented for nonlinear vibratory systems. The reference [10] is a useful source for this topic. Nonlinear vibration control has been used alongside energy harvesting in [11]. Hybrid time-domain and spatial filtering nonlinear damping strategy for efficient broadband vibration control is developed and discussed in [12]. For nonlinear vibrations, a series of works are developed and implemented [13-15].

A set of nonlinearities are caused by the nonlinear geometry of the system, for those, Method of Multiple Scales are used [16-20]. Cantilever beams are typically vibrated nonlinearly when the amplitudes are high. Some useful references of these techniques are found at [21-26].

## 4. POSITIVE POSITION FEEDBACK (PPF)

Positive Position Feedback (PPF) is undoubtedly the most famous technique for vibration control in resonant frequencies. PPF control has been extensively used in space structures vibration control. In fact, PPF was first introduced for this application and was designed to use piezoelectric actuators/sensors [27, 28]. Vibration control of space structures has been a challenging problem since the beginning of space travel. There are several studies on active vibration control of space structures where collocated control methods are widely used. Recently, PPF control has overshadowed other collocated methods enhanced by some other approaches such as adaptive control [29-33]. PPF has been modified in order to have a higher level of suppression [34-36]. Direct velocity feedback and acceleration feedback have also been used by several researchers for the vibration control of space structures [37-45]. Specifically, acceleration feedback control has been used for the control of the self-mobile space manipulator [46].

Additionally, active noise and vibration control of flexible structures by means of smart materials, especially piezoelectric patches, is of interest of many researchers. Application of piezoelectric actuators and shape memory alloys in vibration control is increasing in many research areas from micro-scale actuators in atomic force microscopes to active vibration control of aircraft bodies [47-50]. Direct Velocity Feedback (DVL) and PPF have been experimentally used to control the vibration of a micro-actuator for hard disk drives [51]. Vibrations in an aircraft or aerospace structure may appear due to various issues, and there are different methods to control the vibrations. Active vibration control also has been used for space structures, such as the Solar Array Flight Experiment (SAFE) structure during its deployment [52]. Two of the most recent approaches that are based on PPF are presented in [53, 54].

## 5. VIBRATION CONTROL IN MEMS SYSTEMS

The vibration control of Micro-Electro-Mechanical structures is an interesting and challenging research area that is extensively applicable in micro-mass measurement, micro-sensors, and micro-mirror control. One of the important MEMS devices is micro-gyroscope. Micro-gyroscopes provide a low cost inertial measurement of rotation rate by sensing the Coriolis force, the study of their control is essential [55-57]. An integral part of most MEMS devices is a micro-cantilever. They are the sensing device in micro-biosensors, micro-mass sensors, and in the Atomic Force Microscope. Piezoelectric materials are extensively used in novel studies and industries, especially aerospace, and in wide and various applications in both

macro- and micro-technologies. Because of their small size and light weight, they have been extensively used in aircraft and aerospace structures for active vibration control alongside collocated control methods. They have been used by research institutes such as McDonnell Douglas Aerospace in Huntington Beach, California [58-59]. Acceleration feedback control is not as popular as velocity or position feedback but is still used for many aircraft vibration control applications [60].

Scanning Probe Microscopes (SPM) was initially designed to capture three dimensional images of nano-scale surfaces; however, today it has many other applications including bio-sensing for cell property measurement, nano-manipulation, and friction measurement. Modeling and calculation of the forces between the SPM tip and the sample is one important part of the measurements. There are two forces that should be measured: the Van der Waals force and the contact force. Contact force identification using the subharmonic resonance of contact mode AFM was studied considering the nonlinear contact force between the tip and a hard sample [61]. In another study, the dynamics of the AFM were investigated in the presence of a nonlinear contact and Van der Waals force; however, the micro-cantilever beam was considered to act linearly [62]. However, most of the studies considered forces to act linearly [63]. Nonlinear behavior of non-contact tapping-mode AFM was studied in the presence of the Van der Waals force to study the stability of the system [64]. The dynamic-coupling effect associated with using an iterative control and positive velocity and position feedback control of piezoelectric tube scanners has been studied [65]. An iterative control approach has also been used for high-speed force-distance measurements using AFM [66].

## 6. OTHER NOVEL CONTROL TECHNIQUES

As the last section of this paper, some other novel techniques for vibration control are briefed. In a series of works, vibration control using network based methods are implemented. Two of these studies include PPF-based control and an integral resonant method [67-68]. A novel active pneumatic vibration isolator through floor vibration observer has been used for robust control in [69]. In another study in [70], a self-sensing and actuating method is used. A two-degree-of-freedom active vibration control of a prototyped smart rotor has been investigated [71]. Flatness-based active vibration control for piezoelectric actuators is studied in [72]. For more unconventional techniques that are designed for a variety of systems, see [73-77].

## 7. CONCLUSION

Here in this paper, we presented a comprehensive review of vibration control techniques, with exclusive focus on recently published methods. These techniques were divided into categories of integral based methods, nonlinear techniques, PPF-based methods, vibration control in MEMS systems and some other unconventional methods. Each of the mentioned techniques are specifically designed for a special vibration control case. The controller designed must first understand his problem very well, and then start selecting the controller from available methods. However, it is highly recommended that the technique is modified for the requirements of that specific vibration control problem.